\documentstyle[12pt]{article}
\begin{document}
\begin{center}
{\Large\bf Cavity Field Reconstruction at Finite Temperature}
\end{center}
\begin{center}
H. MOYA-CESSA,$^{1,3}$  A. VIDIELLA-BARRANCO,$^2$ P. TOMBESI,$^1$, 
and J.A. ROVERSI$^2$
\end{center}
\begin{center}
$^1$ Dipartimento di Matematica e Fisica, Universit\`a di Camerino, and INFM 
Unit\`a Camerino I-62032, Camerino (MC), Italy \\
$^2$ Instituto de F\'\i sica ``Gleb Wataghin'',
Universidade Estadual de Campinas,
\\ 13083-970   Campinas  SP  Brazil \\
$^3$ INAOE, Coordinaci\'on de Optica, 
Apdo. Postal 51 y 216,\\ 72000 Puebla, Pue., Mexico.
\end{center}
\begin{abstract}
We present a scheme to reconstruct the quantum state of a 
field prepared inside a lossy cavity at finite temperature.
Quantum coherences are normally destroyed by
the interaction with an environment,
but we show that it is possible to recover complete information
about the initial state (before interaction with its environment), making possible 
to reconstruct any $s$-parametrized quasiprobability distribution, in particular, 
the Wigner function.
\end{abstract}
\newpage
\section{Introduction}
Recently there have been proposals to reconstruct the quantum state of 
electromagnetic fields inside cavities \cite{Lutt,Moy}. The reconstruction of non-
classical states is a central topic in quantum optics and related fields, 
and there have been a number of proposals to achieve it (see for instance 
\cite{ulf}).
In fact, the full reconstruction of nonclassical field states \cite{smi} as well as of 
(motional) states of an ion \cite {win} have been experimentally accomplished. 
The reconstruction is normally achieved through a finite set of either field 
homodyne measurements, or selective measurements of atomic states \cite{Lutt}.

However, the presence of noise and dissipation has normally 
destructive effects. In fact, the reconstruction schemes themeselves also indicate 
loss of coherence in quantum systems \cite{win}. Schemes for compensation of 
losses have already been proposed \cite {Kiss} and the relation between losses and 
$s$-parametrized quasiprobabilities has been pointed out in ref \cite{Ulf2}. A method of
reconstruction of the Wigner function that takes into account losses (at $T=0$)
has also been presented \cite{Moy}.

We consider here a single mode high-$Q$ cavity where we suppose that a 
nonclassical field
is prepared. The first step of our method consists in driving the generated state
by a coherent pulse. The reconstruction of the field is done after turning-off the 
driving
field, i.e. at a time when the cavity field has interacted with its 
environment at finite temperature.
We show that by measuring the density matrix diagonal 
elements and properly weighting them, 
we can obtain directly the Wigner function even at $T \neq 0$. 
We should remark that to know a state,
one has to have information about all the density matrix elements (diagonal and 
off-diagonal),
however, with the method presented here (see also \cite{Moy}), it is only necessary to have 
information about diagonal matrix elements. 

\section{Master equation and its solution}

The master equation in the interaction picture for the reduced density operator
$\hat{\rho}$ relative to a driven cavity mode, taking into account cavity 
losses at non-zero temperature and under the Born-Markov approximation is 
given by \cite {loui} (in a frame rotating at the field frequency $\omega$)

\begin{equation}
\frac{\partial \hat{\rho}}{\partial t}
=(\hat{\cal R}+\hat{\cal L})\hat{\rho},
\label{1}
\end{equation} 
where 
\begin{equation}
\hat{\cal L}\hat{\rho} = (\hat{\cal L}_1 +\hat{\cal L}_2)\hat{\rho}
\label{2}
\end{equation}
with
\begin{equation}
\hat{\cal L}_1\hat{\rho}=\frac{{\Large \gamma}(\bar{n}+1)}{2} \left(
           2\hat{a} \hat{\rho} \hat{a}^\dagger 
             - \hat{a}^\dagger \hat{a} \hat{\rho} - \hat{\rho} \hat{a}^\dagger
             \hat{a}\right), \ \ \ \
\hat{\cal L}_2\hat{\rho}=\frac{\gamma\bar{n}}{2} \left(
           2\hat{a}^\dagger \hat{\rho} \hat{a} 
             - \hat{a} \hat{a}^\dagger \hat{\rho} - \hat{\rho} \hat{a}
             \hat{a}^\dagger\right),
\label{3}
\end{equation}
and
\begin{equation}
\hat{\cal R}\hat{\rho}=-\frac{i}{\hbar} [\hat{H},\hat{\rho}],
\label{4}
\end{equation}
where
\begin{equation}
\hat{H} = i\hbar \left(\alpha ^*\hat{a}-\alpha \hat{a}^\dagger\right).
\label{5}
\end{equation}    
$\hat{a}$ and $\hat{a}^\dagger$ are the annihilation and
creation operators, $\gamma$ the (cavity) decay constant, $\bar{n}$ is the mean
number of thermal photons  and $\alpha$
the amplitude of the driving field. 

\subsection{Displacing the field}

The formal solution to Eq. (\ref{1}) is given by (see for instance \cite {arevalo})

\begin{equation}
\hat{\rho}(t) = \exp\left[(\hat{\cal R}+\hat{\cal L})t\right]\hat{\rho}(0) .
\label{sol}
\end{equation}

It is not difficult to show that Eq. (\ref{sol}) can be factorized in the product of two 
exponentials, one containing the reservoir (super) operators and the other the 
interaction
(\ref{5}), the latter one yielding an effective displacement on the initial field. In 
order to 
show this we calculate the commutator

\begin{equation}
[\hat{\cal R},\hat {\cal L}]\hat{\rho} = \frac{\gamma}{2} 
\hat{\cal R}\hat{\rho},
\label{6}
\end{equation}
which allows the factorization

\begin{equation}
\hat{\rho}(t) =\exp(\hat{\cal L}t)\exp\left[-\frac{2\hat{\cal R}}{\gamma}
          (1-e^{\gamma t/2})\right]\hat{\rho}(0). 
\label{7}
\end{equation}

After driving the initial field during a time $t$, the resulting field density
operator will read

\begin{equation}
\hat{\rho}(t) = e^{\hat{\cal L}t} \hat{\rho}_{\beta}(0),
\label{8}
\end{equation}
where
\begin{equation}
\hat{\rho}_{\beta}(0) = \hat{D}^\dagger(\beta)\hat{\rho}(0)\hat{D}(\beta),
\label{9}
\end{equation}
with the effective (displacing) amplitude

\begin{equation}
\beta = -2\alpha \frac{1-e^{\gamma t/2}}{\gamma}.
\label{10}
\end{equation}

\subsection{Solution to the Master Equation at finite temperature}

We now obtain the density matrix (\ref{4}), by defining 
\cite{arevalo} 
\begin{equation}
\hat{J}_-\hat{\rho}=\hat{a}\hat{\rho}\hat{a}^\dagger, \ \ \ \
\hat{J}_+\hat{\rho}=\hat{a}^\dagger \hat{\rho} \hat{a}, \ \ \ \ 
\hat{J}_3\hat{\rho}=\hat{a}^\dagger\hat{a}\hat{\rho}+\hat{\rho}\hat{a}^
\dagger\hat{a}+\hat{\rho},
\label{11}
\end{equation}
where the superoperators $\hat{J}_-$, $\hat{J}_+$ and $\hat{J}_3$ obey the 
commutation 
relations $[\hat{J}_-,\hat{J}_+]\hat{\rho}=\hat{J}_3\hat{\rho}$ and
$[\hat{J}_3,\hat{J}_\pm]\hat{\rho}=\pm 2\hat{J}_\pm\hat{\rho}$,
which can be written as
\begin{equation}
\hat{\rho}(t) = e^{\frac{\gamma t}{2}} e^{\Gamma_{\bar{n}}(t)\hat{J}_+}
\left[\frac{e^{-\gamma t/2}}{1+N_t}\right]^{\hat{J}_3}
e^{\Gamma_{\bar{n}+1}(t)\hat{J}_-} \hat{\rho}_{\beta}(0),
\label{12}
\end{equation}
and we have defined
\begin{equation}
\Gamma_{\bar{n}}(t)=\frac{\bar{n}(1-e^{-\gamma t})}{1+N_t}, \ \ \ \
\Gamma_{\bar{n}+1}(t)=\frac{(\bar{n}+1)(1-e^{-\gamma t})}{1+N_t},
\label{13}
\end{equation}
with $N_t=\bar{n}(1-e^{-\gamma t})$.

\section{The reconstruction method}

We now calculate the diagonal density matrix elements $<m|\hat{\rho}(t)|m>$ 
from
(\ref{12}) 

\begin{equation}
\langle m|\hat{\rho}_\beta (t)|m\rangle=\frac{1}{1+N_t}
\sum^\infty_{k=0}\sum^\infty_{n=0}
\left(\begin{array}{c} k \\ n \end{array} \right)
\left(\begin{array}{c} m \\ n \end{array}\right)
[\Gamma_{\bar{n}}(t)]^{m-n}[\Gamma_{\bar{n}+1}(t)]^{k-n}
\frac{e^{-n\gamma t}}{[1+N_t]^{2n}}
\langle k|\hat{\rho}_\beta(0)|k\rangle.
\label{14}
\end{equation}
Multipliying the expression above by powers of the weight function $\chi_s$, where 
\begin{equation}
\chi_s=\frac{\frac{s+1}{s-1}-\Gamma_{\bar{n}+1}(t)}{\frac{e^{-\gamma 
t}}{[1+N_t]^2}+
\Gamma_{\bar{n}}(t)\left(\frac{s+1}{s-1}-\Gamma_{\bar{n}+1}(t)\right) }
\end{equation}
and adding over $m$ we obtain

\begin{equation}
F(\beta;s) =\sum^\infty_{m=0}\chi^m_s\langle m|\hat{\rho}_\beta 
(t)|m\rangle
=\frac{1}{[1+N_t][1-\chi_s\Gamma_{\bar{n}}(t)]}
\sum^\infty_{k=0}\left(\frac{s+1}{s-1}\right)^k
\langle k|\hat{\rho}_\beta(0)|k\rangle,
\label{16}
\end{equation}
where we have used the fact that \cite{Abra}
\begin{equation}
\sum^\infty_{m=0}   \left(\begin{array}{c} m \\ n \end{array} \right) x^m= 
\frac{x^n}{[1-x]^{n+1}}.
\end{equation}

If we now multyply $F(\beta;s)$ by the quantity
\begin{equation}
-\frac{2[1+N_t][1-\chi_s\Gamma_{\bar{n}}(t)]}{\pi(s-1)},
\end{equation}
we finally obtain
\begin{equation}
W(\beta;s)= -\frac{2}{\pi(s-1)}\sum^\infty_{k=0}\left(\frac{s+1}{s-1}\right)^k
\langle k|\hat{\rho}_\beta(0)|k\rangle,
\end{equation}
which is the $s$-parametrized quasiprobability distribution \cite{moya}.

Therefore, by measuring the diagonal elements of the evolved field 
density matrix, Eq. (\ref{14}), we may obtain complete information on the
initial state.

We should remark that in this case (thermal environment), more information
on the $P_m(t)=<m|\hat{\rho}(t)|m>$ is required than in the zero temperature
case. Nevertheless, it is possible 
to find a weight function that allows the reconstruction of the initial
field.

\subsection{Measuring the photon distribution}

For completness, we suggest a way to measure the photon number distribution 
$P_m(\beta,t)$ of Eq. (\ref{16}). It is not difficult to show that the atomic 
inversion for the case of a three-level atom in a cascade configuration with the 
upper and the lower levels having the same parity and satisfaying the two-photon 
resonance condition is given by (see for instance \cite{Dantsker})

\begin{equation}
W(\beta;t+\tau) \cong  \sum^\infty_{m=0} P_m(\beta,t) 
\cos([2m+3]\lambda\tau) ,
\label{18}
\end{equation}
where  $\lambda$ is the atom-field coupling constant. In order to obtain $P_m$ 
from a family of measured population inversions, we invert the Fourier series in 
Eq. (\ref{18}), or
\begin{equation}
P_m(\beta,t)=\frac{2\lambda}{\pi}\int_0^{\tau_{max}} d\tau W(t+\tau)   
\cos([2m+3]\lambda\tau) .
\label{19}
\end{equation}
We need a maximum interaction time $\tau_{max}=\pi/\lambda$ much 
shorter than the cavity decay time, which implies we must be in the strong-coupling 
regime, i.e. $\lambda>>\gamma$.

\section{Conclusions}

We have presented a method to reconstruct 
the Wigner function (and in general any quasiprobability distribution) of an 
initial nonclassical state at times
when the field would have normally lost its quantum coherence because of the 
interaction with an environment at finite temperature. This is an extension of 
our previous work \cite{Moy} where we considered the interaction with the 
environment at zero temperature.
The crucial point of our approach is the 
apropriate weighting of the evolved (driven and decayed)
photon number distribution.
Driving the initial
field immediately after preparation, is not only useful for
covering a region in phase space but also makes possible, together with the weight function, 
$\chi_s$, to store quantum coherences
in the diagonal elements of the time evolved density
matrix. 
We have shown here that it is possible to find a
weight function that allows to reconstruct the initial field even 
at finite temperature, and this is the main result of our paper.
Similar conclusions may be reached by employing the method of generating functions
\cite {Dodonov}.
The possibility of reconstructing quantum 
states when the system interacts with an environment maybe relevant for 
applications in quantum computing. Loss of coherence for such interactions is 
likely to occur in such devices, and our method could be used, for instance, as a 
scheme to refresh the state of a quantum computer \cite{Ekert} in order to 
minimize the destructive action of a hot environment.

{\large\bf Acknowledgments}

This work was partially supported by CONACYT (Mexico), FAPESP and CNPq 
(Brazil).

%
%

%
%

\end{document}